\documentclass[aps,prl,epsfigure,twocolumn,superscriptaddress]{revtex4-1}
\usepackage{dcolumn} 
\usepackage{bm} 
\usepackage{graphicx}
\usepackage{amsmath}    
\usepackage{latexsym}
\usepackage{amsfonts}   
\usepackage{amssymb}
\usepackage{array}      
\usepackage{epsfig}
\usepackage{txfonts}
\usepackage{color}
\usepackage[colorlinks=true,linkcolor=blue,urlcolor=blue,citecolor=blue]{hyperref}
\usepackage{hyperref}
\usepackage{enumerate}
\usepackage{bbold}
\usepackage{cleveref}
\usepackage{appendix}
\DeclareMathOperator{\sign}{sgn}

\newcommand{\hhu}{Institute for Theoretical Physics II: Soft Matter, Heinrich-Heine-Universit\"at D\"usseldorf, Universit\"atsstr. 1, 40225 D\"usseldorf, Germany}
\newcommand{\dlrai}{German Aerospace Center (DLR), Institute for AI Safety and Security, Wilhelm‐Runge‐Str.~10, 89081 Ulm, Germany}
\newcommand{\dlrmp}{German Aerospace Center (DLR), Institute for Material Physics in Space, 51170 Köln, Germany}

\begin{document}
\title{Universal Hyperuniformity in Active Field Theories}
\author{Yuanjian Zheng} 	
\affiliation{\hhu}
\author{Michael A. Klatt}
\affiliation{\hhu}
\affiliation{\dlrai}
\affiliation{\dlrmp}
\author{Hartmut L\"owen}
\affiliation{\hhu}

\begin{abstract}     
	
We show that dry scalar-order active field theories (AFTs) are
universally hyperuniform, i.e., density fluctuations are anomalously
suppressed in the long-time limit regardless of the integrability or
functional form of the active contributions up to third order in
gradient terms. These AFTs include Active model B, Active model B+, and
effective Cahn-Hilliard models. Moreover, density variances and spectral
densities are virtually indistinguishable from that of passive
phase-separated hyperuniform fields. Higher moments of the density
fluctuations, however, reveal activity-dependent higher-order
correlations that are not captured by conventional two-point measures
that characterize hyperuniformity.
	
\end{abstract}	
\maketitle

Active systems are driven out-of-equilibrium by
the continuous consumption and dissipation of energy
\cite{MarchettiSimha2013, BechingerVolpe2016, Ramaswamy2010}. These include all biological living matter and a growing number of synthetic
systems realized across various length scales
\cite{TheurkauffBocquet2012, ScholzLoewen2018, YuZhang2018}. Entropy in
active matter is produced locally \cite{FodorWijland2016,
MandaldeWeese2017}, which results in dynamics and collective behavior
that differs significantly from passive systems, constituting robust
frameworks for self assembly of complex structures or the emergence of
novel material properties \cite{ScheibnerVitelli2020, ZottlStark2016,
SchwarzLinek2012}. In particular, active matter composed of
self-propelled particles can separate into co-existing
density-differentiated phases in the absence of any microscopic
attractive or alignment interactions \cite{CatesTailleur2015,
FilyMarchetti2012, RednerBaskaran2013}. This remarkable behavior
known as ``Motility Induced Phase Separation" (MIPS)
\cite{CatesTailleur2015}, has since become a hallmark of active
phenomena. To understand the underlying mechanisms for symmetry breaking in MIPS, several continuum models have been proposed to represent the dynamics in
the hydrodynamic limit of the  density field $\phi(\bm{r},t)$. These
{active field theories} (AFTs) \cite{SpeckLoewen2014,
WittkowskiCates2014, TjhungCates2018, Cates2019} come in the form of
extending a (passive) field theory for Brownian motion - \emph{model-B}
\cite{HohenbergHalperin1977}, by introducing an additional scalar-order
active term $ g_\mathcal{X} \mathcal{A}_\mathcal{X}[\phi]$ to the
\emph{Cahn-Hilliard} (CH) equation that models passive phase separation
in the study of binary fluids \cite{CahnHilliard1958}.
\begin{equation}
  \partial_{t} \phi (\bm{r},t)= D \nabla^2 \mu[\phi] + g_{\mathcal{X}} \mathcal{A}_{\mathcal{X}}[\phi]
  \label{eq:eom}
\end{equation}
where the chemical potential -- $\mu[\phi]=f'[\phi]-\gamma_{a} \nabla^2
\phi $ is the functional derivative of the equilibrium free-energy
\begin{equation}
  \mathcal{F}[\phi]=\int d\bm{r} \left\{  f[\phi] +\frac{\gamma_{a}}{2} \lvert \nabla \phi \rvert^2  \right\}
  \label{eq:free_energy}
\end{equation}
that is locally bi-stable for phenomenological reasons \cite{Cates2019}
and written in Ginzburg-Landau form -- $f[\phi]=\frac{1}{4}(\phi^2-1)^2$
where $D$ and $\gamma_a$ control diffusivity and width of the segregated
phases at long times respectively.

\begin{figure*}[t!]	
  \includegraphics[width=0.8\textwidth]{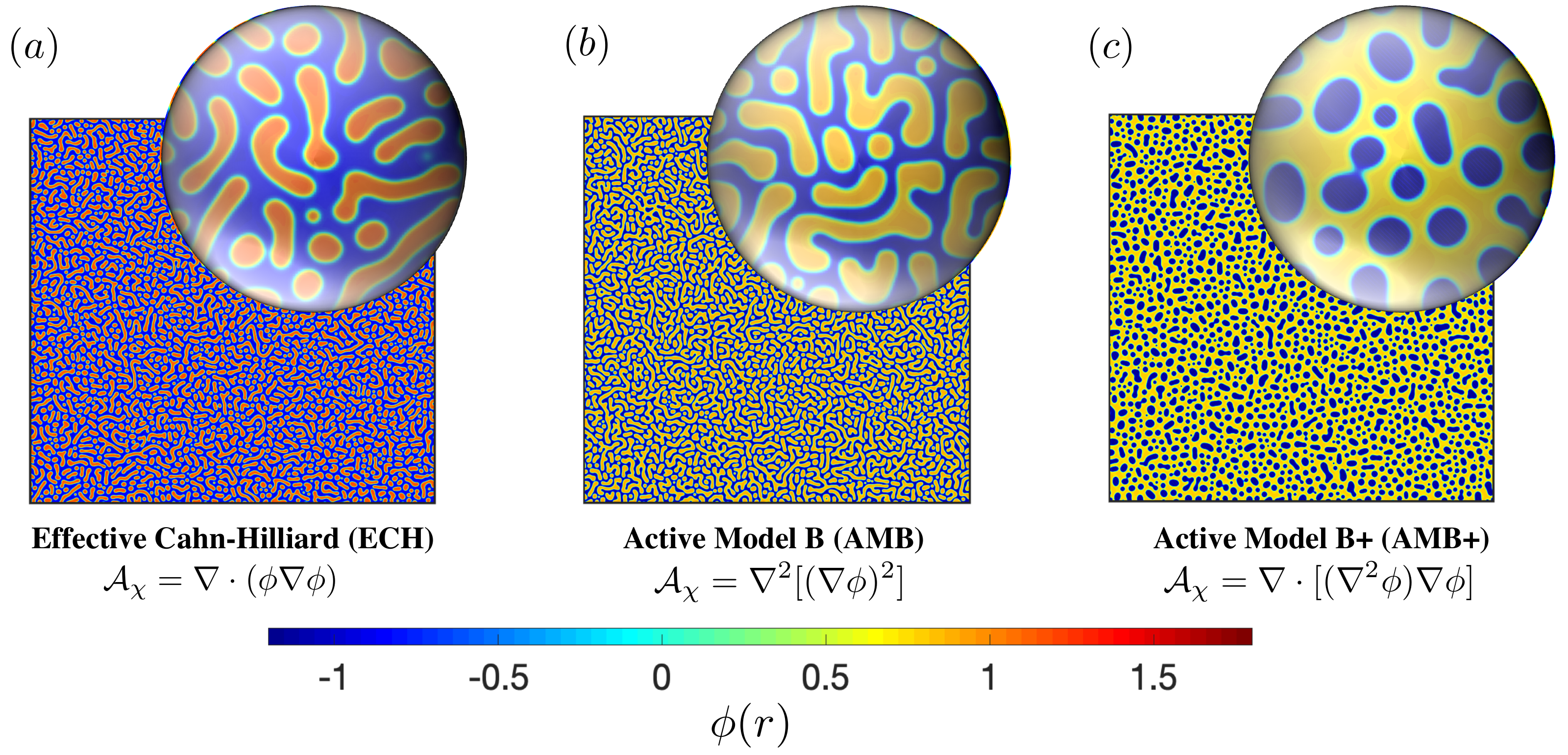}		
  \caption{$\phi_{t=\tau}(\bm{r})$ of scalar-order AFTs for various
  active terms -- $ g_{\mathcal{X}}\mathcal{A}_{\mathcal{X}}[\phi(
  \bm{r})]$. {\bf (a)} $g_{ECH}/D=0.2$, { \bf (b)} $g_{AMB}/D=-0.5$, and
  { \bf (c)} $g_{AMB+}/D =2$;
  for all samples,
  $L=1024, D=0.01, \gamma_a =2$, and $\tau =10^3\gamma_a/2D$.}
  \label{fig:configuration}
\end{figure*}

The functional form of $\mathcal{A}_{\mathcal{X}}[\phi]$ has previously
been proposed by coarse-graining the microscopic equations of motion
such as in the \emph{effective} Cahn-Hilliard model --
$\mathcal{A}_{ECH}=\nabla \cdot (\phi \nabla \phi)$
\cite{SpeckLoewen2014}. However, this form of
$\mathcal{A}_{\mathcal{X}}[\phi]$ remains integrable in that the
dynamics can still be written in an effective free energy form through
modification of (\ref{eq:free_energy}). Hence, alternate representations
for $\mathcal{A}_{\mathcal{X}}[\phi]$ have subsequently arisen from more
field-theoretic approaches that invoke minimal non-integrable terms that
violate detailed balance, such as active model B -- $\mathcal{A}_{AMB} =
\nabla^2 [(\nabla \phi)^2]$ \cite{WittkowskiCates2014}, or active model
B+ -- $\mathcal{A}_{AMB+} = \nabla \cdot [(\nabla^2 \phi) \nabla \phi]$
\cite{TjhungCates2018}; see Fig.\ref{fig:configuration}.
While these variants produce striking differences in $ \phi(\bm{r},t)$ at
short cluster-sized length scales -- $l_1(t)$, little is
known of their spatial structure at larger scales, where non-trivial long-range correlations may exist in both
AFTs and the underlying atomistic active matter they represent.
Moreover, the degree to which these generic AFTs describe
structural characteristics of agent-based simulations or experiments
that involve non-standard interactions or environments
\cite{ZhangGranick2021, TungCacciuto2016, AndersonFernandezNieves2022, KalzMetzler2023}
is also unknown.

At the same time, recent studies in chiral or circle
micro-swimmers
\cite{LeiCiamarraNi2019,LeiNi2019,Torquato2021,HuangZhang2021,ZhangSnezhko2022,OppenheimerEtAl2022,KurodaMiyazaki2023a}
indicate that certain active systems possess a form of hidden long-range
order known as \emph{hyperuniformity}
\cite{TorquatoStillinger2003,Torquato2018,Torquato2016} in which density
fluctuations are anomalously suppressed even at infinite length scales.
Here, the scalar density field $\phi(\bm{r},t)$ in the
long-time limit is of interest, i.e., the scaling regime where $\phi(\bm{r},t)$ becomes independent of $t$ except for a trivial rescaling by $l_1(t)$. {In other words, our results hold in the thermodynamic limit, for arbitrary times in the scaling regime.} We hence drop the time dependence in our notation hereafter.
For scalar fields, hyperuniformity is characterized by a vanishing spectral density
in the long wavelength limit -- $\lim_{\bm{k} \to 0}
\tilde{\psi}(\bm{k})=0$, where $\tilde{\psi}(\bm{k})$ is the Fourier
transform of the density auto-covariance function  --
$\psi(\bm{r})=\langle (\phi(\bm{r}_i )-\bar{\phi})  (\phi(\bm{r}_j
)-\bar{\phi}) \rangle$ where $\bm{r}=\bm{r}_j-\bm{r}_i $\cite{Torquato2016,MaTorquato2017}, analogous to the static
structure factor $S(\bm{k})= \frac{1}{N}\rvert \sum^N_j e^{-i \bm{k}
\cdot \bm{r_j}} \lvert^2$ defined for point patterns $\{\bm{r_{j} }\}$.
Alternatively, disordered hyperuniformity can also be defined by the
index $\lambda > d$ in the power-law scaling dependence of the variance
-- $ \lim_{R \to \infty}\sigma^2(R) \sim R^{-\lambda}$ for
coarse-grained density $\phi_R=\frac{1}{R^d} \int_{\Omega_R} d^d \bm{r}
\phi(\bm{r})$, where $d$ is the spatial dimension, $R$ the
coarse-graining length scale, and $\Omega_R$ the corresponding observation window.
These results come especially surprising,
given that activity typically induces phase separation and
density inhomogeneities at macroscopic lengths
\cite{CatesTailleur2015,FilyMarchetti2012,RednerBaskaran2013}, while
collective behavior in polar active matter such as flocking, leads
generically to giant number fluctuations
\cite{RamaswamyToner2003,ChateMontagne2006, NarayanMenon2007,
KurodaMiyazaki2023b}; both of which, conceivably, only serve to
\emph{enhance} density fluctuations. {Recent discoveries of hyperuniformity in several hydrodynamic formulations of other active systems (e.g. \cite{MaCates2023,BoltzIhle2024}) add further intrigue to the riddle.} These seemingly paradoxical observations hint at a more fundamental connection between hyperuniformity and activity.

In this Letter, we establish such a connection by showing that dry
scalar-order AFTs are class I hyperuniform (the strongest class
of hyperuniformity). In particular, we show that
$\tilde{\psi}(k)
\sim k^4$ and $\sigma^2(R) \sim R^{-(d+1)}$ for AFTs up to
$\nabla^3\phi$ terms in $\mathcal{A}_{\mathcal{X}}$ ($d=2$)
regardless of the integrability or functional form of the active
contribution. This occurs in spite of stark differences in structure
at shorter, cluster-sized length scales -- $l_1$, which are comparable in scale to the hyperuniform length -- $\xi_h$
that characterizes the onset of fluctuation suppression. Additionally,
while AFTs are indistinguishably hyperuniform, hidden correlations that
lie beyond $\xi_h$ exists, and is revealed through
examining central moments of the coarse-grained probability density -- $P(\phi_R,
R)$. These features differentiate AFTs and persists even at $R >\xi_h$, and may thus be used to
establish correspondence with fluctuation behavior in atomistic active
matter that involve novel interactions, geometry or dynamics
\cite{ZhangGranick2021,TungCacciuto2016,AndersonFernandezNieves2022,KalzMetzler2023} .

We begin by first solving equations (\ref{eq:eom}) \&
(\ref{eq:free_energy}) for the various AFTs in 2D {using a finite (central) difference method with step size of unit time \cite{numerical}} --  (I) Effective Cahn-Hilliard model: $\mathcal{A}_{ECH}=\nabla \cdot (\phi \nabla \phi)$, $g_{ECH}/D=0.2$ (II) Active model-B: $\mathcal{A}_{AMB} = \nabla^2 [(\nabla \phi)^2]$, $g_{AMB}/D=-0.5$
and (III) Active model-B+: $\mathcal{A}_{AMB+} = \nabla \cdot [(\nabla^2 \phi) \nabla
\phi]$, $g_{AMB+}/D=2.0$. Collectively, these systems
represent active contributions up to $3^{rd}$ order in gradient terms of
$\phi(\bm{r})$ exhaustively and include variants that are not
physically motivated by microscopic equations of motion
\cite{Cates2019}. The parameters governing passive diffusive behavior in
(\ref{eq:eom}) are identically fixed for all models considered
($D=0.01$, $\gamma_a=2$), while the initial field $\phi(\bm{r},t=0)$
deviates locally from zero by random displacements drawn from a uniform
distribution bounded by $\pm 10^{-4}$. This choice corresponds to what is known
in binary fluids as the ``critical quench" \cite{MaTorquato2017}
condition. The system is integrated on a $L \times L$ grid for total
time of $\tau=10^3 \gamma_a/ 2D$, where the lattice spacing is the unit of
length. Note that long-range order in $\phi(\bm{r},t)$ is convergent in
$t \to \infty$, even in the absence of fixed points to the CH dynamics
\cite{MaTorquato2017}.  Examples of $\phi_{t=\tau}(\bm{r})$ displaying
activity-dependent variability in structure are shown in
Fig.\ref{fig:configuration} for $g_{\mathcal{X}}$ values chosen to
accentuate structural variability. In particular, observe the
``Bubbly" phase separation \cite{TjhungCates2018}  and 
inversion in relative density of the fully-connected phase for  $\mathcal{A}_{AMB+} $ .
\begin{figure}[t!]	
  \includegraphics[width=0.7\columnwidth]{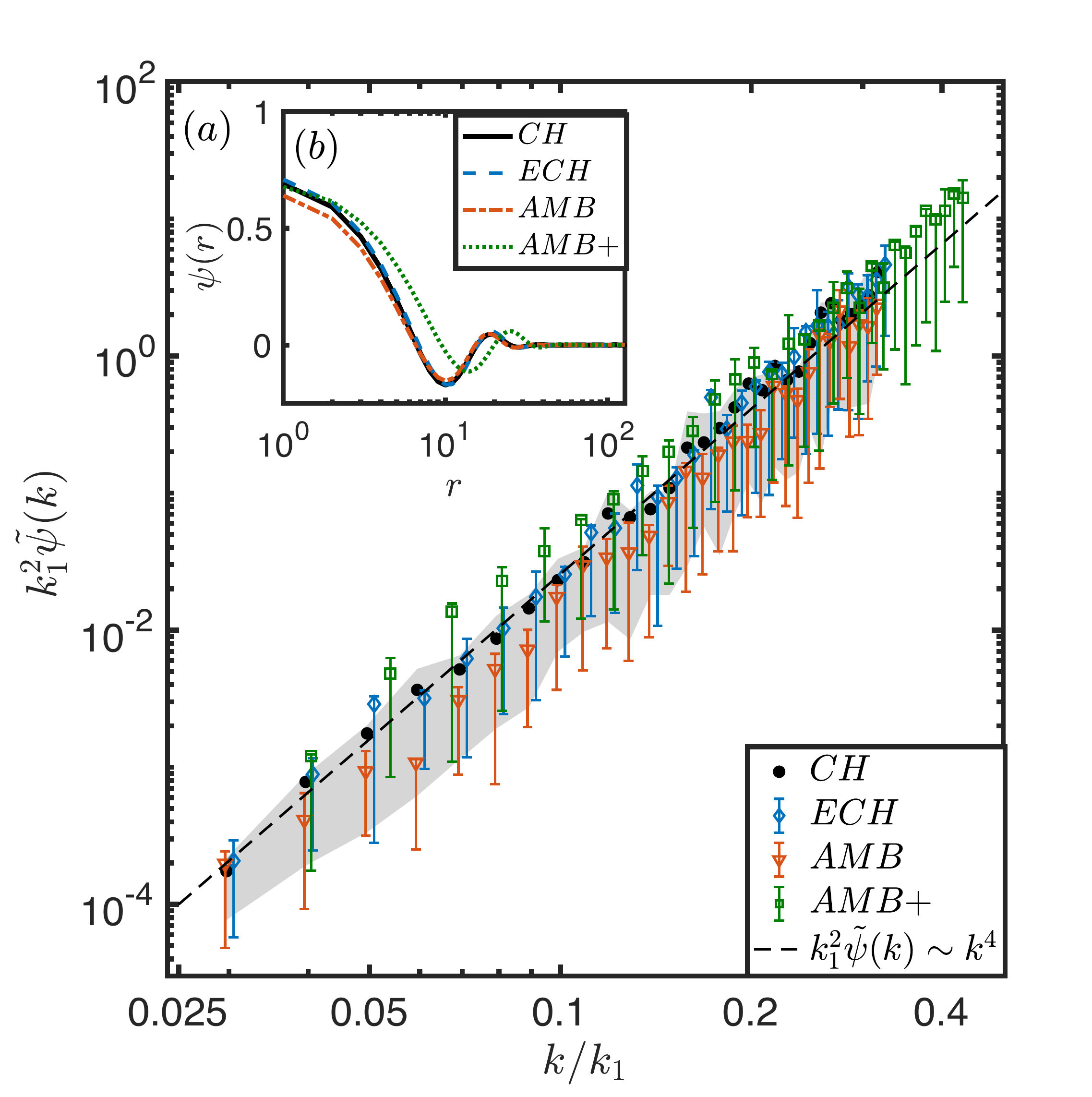}
  \caption{{\bf (a)}  $k_1^2\tilde{\psi}(k)$ and {\bf (b)} $\psi(r)$
  for [{\bf Black}] \emph{passive} model-B (CH), and various AFTs: {\bf
  [Blue]} $g_{ECH}/D=0.2$, {\bf [Red]} $g_{AMB}/D=-0.5$, and {\bf
  [Green]} $g_{AMB+}/D =2.0$.  Colored-symbols in {\bf(a)} indicate mean
  values of $k_1^2\tilde{\psi}(k)$ obtained from $N_s=50$ realizations of $\phi_{t=\tau}(\bm{r})$. The grey-shaded area and
  error bars represent respectively, sample-to-sample variation for
  model-B and corresponding AFTs, characterized by the $1^{st} $and
  $3^{rd}$ quartiles of sample values of $k_1^2 
  \tilde{\psi}(k)$. $k_1=2\pi/l_1$
  characterizes structure at cluster length -- $l_1= \arg \min_r
  \psi(r)$, where peak anti-correlation in $\psi(r)$ occurs in {\bf
  (b)}. The black-dashed line in {\bf (a)} represents the 
  fit to $\tilde{\psi}(k) \sim k^4$ for model-B.}
  \label{fig:structure}
\end{figure}
\begin{figure*}[htb!]	
  \includegraphics[width=0.9\textwidth]{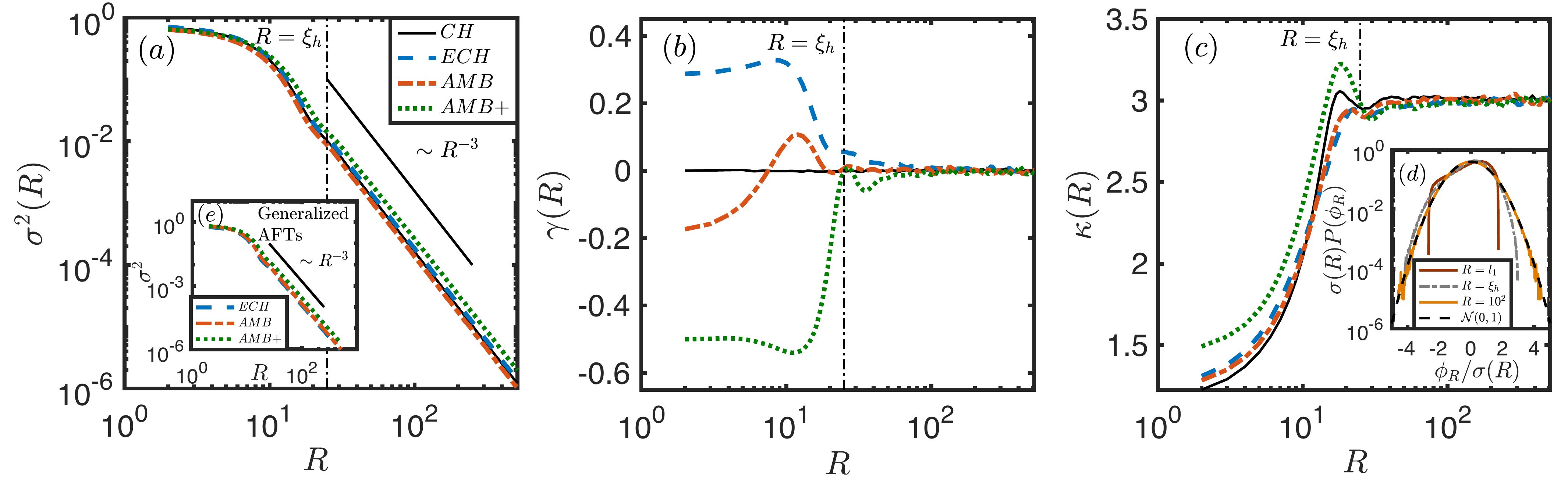}
  \caption{ {\bf(a)} Variance -- $\sigma^2(R)$ {\bf (b)} skewness --
  $\gamma^2(R)$ and {\bf(c)} kurtosis  -- $\kappa^2(R)$ of  $P(\phi_R)$
  . ${\bf (a)}$ Universal suppression of density fluctuations
  $\sigma^2\sim R^{-3}$ indicates that AFTs are indistinguishably
  class I hyperuniform. $\xi_h$(vertical dot-dashed lines) characterizes
  the on-set of fluctuation suppression in $\sigma^2(R)$ and are scale
  comparable for all models considered. {\bf (b, c)} Deviations in the
  higher moment  from their respective Gaussian values
  ($\gamma_{\mathcal{N}}=0$, $\kappa_{\mathcal{N}}=3$) at $R > \xi_h$
  indicates the presence of hidden correlations beyond what is revealed by two-point functions. {Inset {\bf (d)} Convergence of $\sigma(R) P(\phi_R)$ to the standard Gaussian -- $\mathcal{N}(0,1)$
  (Black-dashed line) in the limit of $R \to \infty$ for $g_{AMB+}/D=2$. Inset {\bf (e)} Generalized AFTs for $n=2$ preserves universal hyperuniformity.} Statistics are obtained from $ 10^5$ $R \times R$ windows, for each of $N_s=50$ realizations.
  }
  \label{fig:moments}
\end{figure*} 

The spectral density for various AFTs are shown
in Fig.\ref{fig:structure}(a) to vanish in the limit of $ k \to
0$ as power-laws -- $\tilde{\psi}(k) \sim k^\alpha$. Moreover,
$\tilde{\psi}(k)$ for all AFTs considered, are within sample
variability, indistinguishable from that of passive model-B
($g_{\mathcal{X}}=0$) (Black-Circles)
\cite{CahnHilliard1958} which is known to be class I hyperuniform
($\alpha_{CH}=4$ for $d=2$) \cite{MaTorquato2017,
WilkenSaleh2023}. {Least-square free fits \cite{free_fit} to the data are consistent with $\alpha_{\mathcal{X}}=4$ and comparable to $\alpha_{CH}$ for all $\mathcal{X}$}. In fact, constrained fits where $\alpha=4$
is fixed, yield better fits in some instances, from a
reduced-$\chi^{2}$ measure perspective, indicating that class I
hyperuniformity ($\alpha > 1$) is universal across AFTs (See SI). Hyperuniformity is further corroborated in
Fig.\ref{fig:structure}(b), by presence of regions where
auto-covariance in real space is negative (i.e $\psi(r) < 0$), since
$\int \psi(\bm{r}) d^d \bm{r}=0$ is a necessary condition for
hyperuniformity \cite{MaTorquato2017}. Note that $l_1 \equiv \arg
\min_{r} \psi(r)$ at which peak anti-correlation occurs \cite{peak_anti} is
used to characterize the cluster length scale and  defines wave-number
$k_1=2\pi/l_1$ used in scaling $\tilde{\psi}(k)$ in
Fig.\ref{fig:structure}(a) \cite{Supp}. {Note that $g_{\mathcal{X}}$
considered in this work span a significant range ($g_{\mathcal{X}}/D \sim
10^{-1} - 10^{0}$) where active strengths $g_{\mathcal{X}}$ are scale comparable to the passive contribution $D$ in (\ref{eq:free_energy}). Hence, these results thus represent systems near the soft upper bound for any realistic scenario of AFTs because the phase separation is driven by the diffusive part, and $g/D \gg 1$ would result in un-physical dynamics. The parameters we have considered thus already includes the regime of strong activity. This broad range suggest that emergence of $\alpha=4$ is universal and cannot be understood as a mere perturbative result to the passive CH model}, which would only be valid in the limit $g_{\mathcal{X}}/ D \to 0$. The $g_{\mathcal{X}}$-dependence
for various AFTs is further detailed in the SI which confirms that
$\tilde{\psi}(k) \sim k^4$ holds across wide ranges of $g_{\mathcal{X}}$.

Universality is also verified through $\sigma^2(R)$ of the coarse-grained density $\phi_R$ shown in  Fig.\ref{fig:moments}(a), which are again indistinguishable from model-B. While fluctuations are enhanced
($ \sigma^2 \sim R^0$) at shorter lengths, a continuous cross-over
occurs at $R \sim \xi_h$, to an anomalous hyperuniform scaling regime
($ \sigma^2 \sim R^{-\lambda}$), that is consistent with class I hyperuniformity ($ \alpha > 1$) and prior numerical results
derived from $\tilde{\psi}(k)$ \cite{Supp}. Note that $\xi_h$ for
various AFTs are similar in scale, and are thus represented  in
Fig.\ref{fig:moments}(a-c) by a single vertical dot-dashed line.
Moreover, $ \xi_h/l_1 \lesssim 2$ for all AFTs considered (See SI), and
are thus comparable to $l_1\sim \gamma_a^{-1/2}$ arising from
short-ranged structural order present in the passive CH model
\cite{CahnHilliard1958}. This broad range suggests that fluctuation suppression
should readily be observable in agent-based models at length scales not significantly larger than respective cluster sizes, which in turn
questions the extent to which existing AFTs correctly describe
long-ranged order of the underlying active systems they purportedly
represent, especially for systems where the cluster length scale remains finite and observable (i.e $l_1 < L$). Here, we remark that emergence of hyperuniformity in AFTs is
unlike the behavior of hyperuniform complex systems associated to
self-organized or driven critical phenomena \cite{HexnerLevin2017,
ZhengPicaCiamarra2020, ZhengPicaCiamarra2021, HexnerLevine2015,
HexnerNagel2018, HexnerLevine2017b, WilkenChaikin2020}. In these near-critical systems, short length scale
density fluctuations are anomalously suppressed below a hyperuniform
length -- $\xi'_h$ and conversely scale like un-correlated fields above
$\xi'_h$. Hence, these systems become hyperuniform only at the critical
point as $\xi_h$ grows and diverges on approach to criticality. This
contrasts with hyperuniformity in AFTs where fluctuations are instead
suppressed above $\xi_h$, and do not involve diverging lengths.

Now, despite the convergence of $\tilde{\psi}(k)$ and $\sigma^2(R)$ that
render AFTs indistinguishable in the long wavelength limit, hidden
higher-order correlations exists that are not captured by
these two-point correlation functions of the coarse grained variable
$\phi_R$. These correlations are often revealed by examining the higher
central moments -- $ \int^{\infty}_{-\infty} d\phi_R \phi^m_R P(\phi_R)
$ of the probability density -- $P(\phi_R)$, where the second moment
($m=2$) or variance -- $\sigma^2(R)$ was used previously to establish
hyperuniformity \cite{ZhengPicaCiamarra2021,TorquatoKlatt2021}. In
Fig.\ref{fig:moments}(b,c) we show skewness ($m=3$) -- $\gamma(R)$ and
kurtosis -- $\kappa(R)$ ($m=4$) for various AFTs as a function of the
coarse-graining length scale $R$. At large $R$, a convergence to their
respective Gaussian limits ($\gamma \to 0$ and $\kappa \to 3$),  in
accordance to the Central Limit Theorem (CLT) is recovered. However, at
shorter length scales, $\gamma$ and $\kappa$ exhibit qualitative
and graded differences in their behavior that is not present in
$\sigma^2(R)$, and are direct imprints of higher-order spatial
correlations of the various AFTs \cite{ZhengPicaCiamarra2021,TorquatoKlatt2021}. For example, the convergence to CLT can be non-monotonic in $R$ and $\sign \gamma(R)$ can take on
intermittent positive or negative values; in contrast,
$\gamma(R)=0$ for passive CH (Black-solid line in
Fig.\ref{fig:moments}(b)), where $P(\phi_R)$ remains symmetric for all
$R$. 

Moreover, these deviations from CLT are sensitive to $\mathcal{A}_{\mathcal{X}}$ and persist even
at $R > \xi_h$ in Fig.\ref{fig:moments}(b,c), which indicates that
$P(\phi_R)$ at $R=\xi_h$ remains highly non-Gaussian. {To examine this behavior,
we show $P(\phi_R)$ for $g_{AMB+}/D=2.0$ in
Fig.\ref{fig:moments}(d)} (See SI for other
$\mathcal{A}_{\mathcal{X}}$), where a non-trivial bi-modal to uni-modal
transition occurs with increasing $R$, and a departure of $P(\phi_{R=l_1})$ and $P(\phi_{R=\xi_h})$ from the standard normal distribution -- $\mathcal{N}(0,1)$ is observed, revealing origins for the negative skew and the platykurtic kurtosis (i.e. $\kappa< 3)$  observed in Fig.\ref{fig:moments}(b,c). In fact, we find that while $\vert \vert \gamma(\xi_h)\vert \vert  < \vert \vert \gamma(l_1)\vert \vert$ for all
$\mathcal{A}_{\mathcal{X}}[\phi]$, distributions remain significantly non-Gaussian
-- $\vert \vert \gamma(\xi_h) /  \gamma(10^2)\vert \vert
\sim\mathcal{O}(1)$ \cite{Supp_gaussian} . Hence, while AFTs are indistinguishably
hyperuniform, $P(\phi_R)$ at $R=\xi_h$ remain
distinguishable and non-Gaussian. {We stress that features in $P(\phi_R)$ cannot simply be associated to a particular AFT simply by a visual inspection of the spatial structure of $\phi(\bm{r})$ in Fig.\ref{fig:configuration} given the oscillatory and highly complicated behavior of the higher moments at various $R$}. These features are representations of extended spatial structure not captured by pair-wise relations in two-point measures such
as $\tilde{\psi}(k)$ and $\sigma^2(R)$ and go beyond simply characterizing $l_1$. They are instead, indicative of many-body correlations at \emph{varying} length scales \cite{TorquatoKlatt2021, ZhengPicaCiamarra2021}, and thus could further serve to assess the extent to
which various AFTs agree  with model-specific
atomistic active matter in simulations or experiments.

\begin{figure}[t!]	
  \includegraphics[width=0.85\columnwidth]{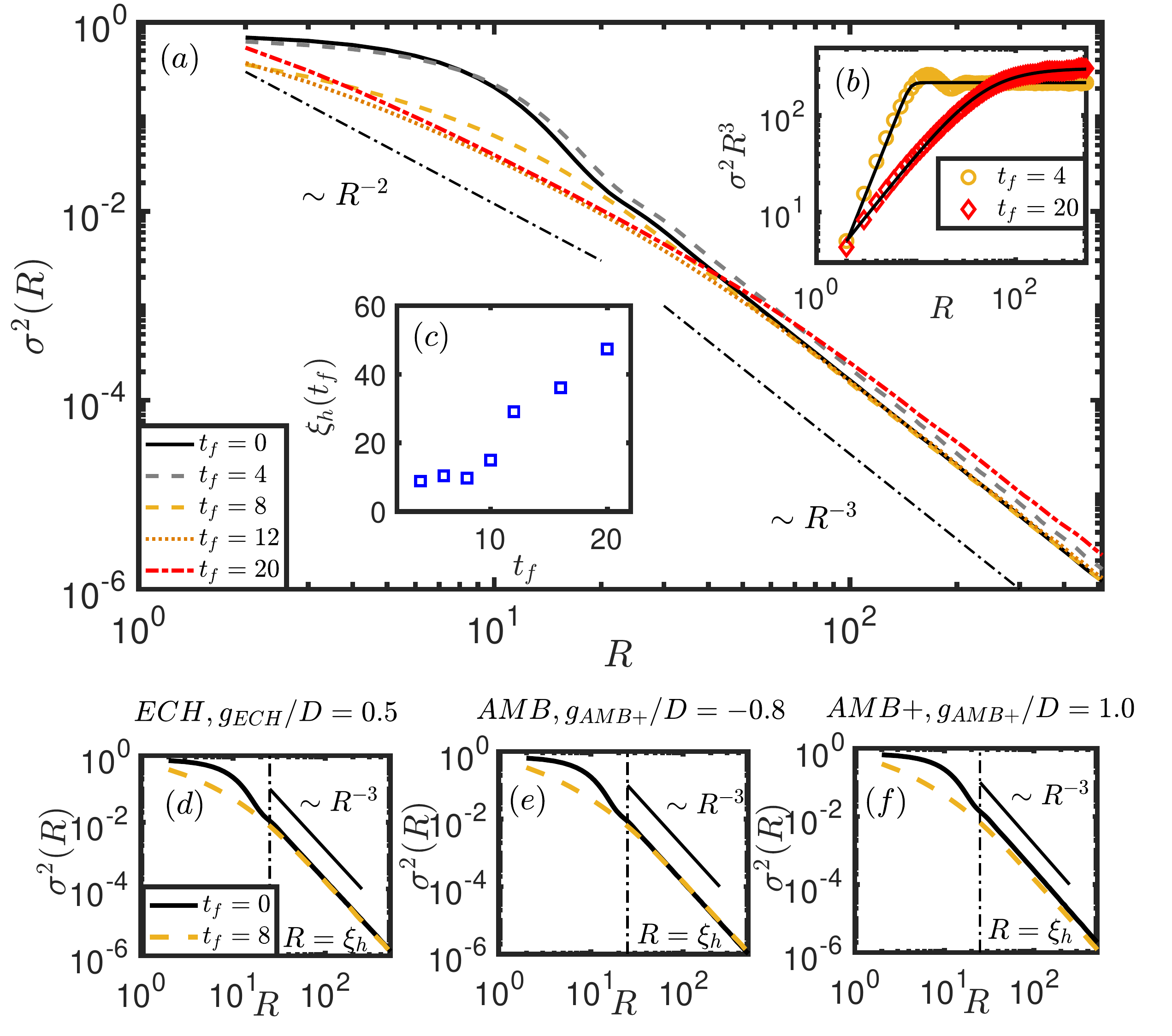}
  \caption{ {{\bf (a)} $\sigma^2(R)$ in the presence of thermal noise at temperature -- $T=D t^2_f/2k_b$, indicate that passive CH remains class I hyperuniform. {\bf (b)} Best-fit (black-solid line) of data to $\sigma^2 R^3 \sim \left[ 1+(R/\xi_h)^{-s}\right]^{-\lambda/s}$ indicate {\bf (c)} a growing $\xi_h$ with increasing $t_f$. Universal class I hyperuniformity in the presence of thermal fluctuations extends to all variants of activity -- {\bf (d)} ECH, {\bf (e)} AMB and {\bf (f)} AMB+.}}
  \label{fig:temeperature}
\end{figure}

{Here, we remark that universal class I hyperuniformity extends even to generalized-AFTs \cite{MaTorquato2017} where $\mathcal{F}[\phi]$(\ref{eq:free_energy}) is not
given by Ginzburg-Landau, but instead assumes a generalized double
well, i.e., $F(n,[\phi])=\int d\bm{r} \left\{ \frac{1}{4n}(\phi^2-1)^{2n}
+\frac{\gamma_a}{2} \vert \nabla \phi\vert^2 \right\}$, where $n$ is a
positive integer and  $F(n,[\phi])$ reverts to the conventional AFTs for
$n=1$; see Fig.\ref{fig:moments}(e) \& SI. Moreover, universal class I hyperuniformity is also robust in the presence of thermal fluctuations, which can be modelled by an additional contribution of $D t_f \nabla \cdot \bm{\eta}$ to $\partial_t \phi$ in (\ref{eq:eom}), where $\langle \eta_i \eta_j \rangle=\delta_{ij}\delta(t-t')$ such that the equilibrium temperature is parameterized -- $T=Dt_f^2/2 k_b$. Conceivably, the presence of dissipation could alter fluctuation suppression or completely destroy long-range order. However, we see in Fig.\ref{fig:temeperature}(a) that hyperuniformity is still preserved for passive model-B. Thermal fluctuations serve instead to progressively destroy short-range order ($ \lim_{t_f \to \infty}\sigma^2(R) \to R^{-2}$) and increases $\xi_h$, see Fig.\ref{fig:temeperature}(b,c), but still always preserves class I scaling, i.e., $\sigma^2(R) \sim R^{-3}$,for its long-range behavior. This robustness of class I hyperuniformity extends to the AFTs (Fig.\ref{fig:temeperature}(d,f)), and can be understood by considering the high-$T$ limit of the dynamics (\ref{eq:eom}) - $\lim_{t_f \to \infty} \phi(\tau) \sim \int_0^\tau dt \nabla \cdot \bm{\eta}$, for which its Fourier transform yields $ \tilde{\phi}(\bm{k}) \sim i\bm{k} \int dt \tilde{\eta}(\bm{k}) \sim \bm{k}$ since $ \int dt \tilde{\eta}(\bm{k})=0$. Hence, 
 $\tilde{\psi}(\bm{k}) \sim \langle 
 \langle \tilde{\phi}_{{\bm{k}}}\tilde{\phi}_{{-\bm{k}}} \rangle \sim \bm{k{^2}}$ and class I hyperuniformity is always preserved in AFTs even in the high-$T$ limit.}

In this work, we show that scalar-order AFTs which describe dry, active
phase-separated matter are class I hyperuniform. Specifically, $\tilde{\psi}(\bm{k})$ vanishes in the  hydrodynamic limit as
$\tilde{\psi}({{k}}) \sim {k}^4$, and density fluctuations are
suppressed $\sigma^2(R) \sim R^{-3}$,  regardless of the
integrability or form of the active contribution --
$\mathcal{A}_{\mathcal{X}}[\phi]$ in (\ref{eq:eom}). This occurs despite
(i) strong variability in short-range structure of $\phi(\bm{r})$, (ii) holds generally for a wide
range of activity, and (ii) persists even in the presence of strong thermal dissipation. This ubiquity in long wavelength fluctuations
across all dry, scalar-order AFTs and their generalizations suggest that
hyperuniformity may be a more generic property of active matter that
goes beyond the context of existing literature on chiral or
circle micro-swimmers \cite{Torquato2021, LeiNi2019, ZhangSnezhko2022,
HuangZhang2021,KurodaMiyazaki2023a}. The robustness of hyperuniformity
in the presence of stochastic noise \cite{HexnerLevin2017,Ikeda2023}, external driving \cite{KurodaIkeda2023,WilkenChaikin2020}, or hydrodynamic
interactions such as Active model H
\cite{TribocchiCates2015}, or in field theories with spatially dependent
diffusion or mobility, are interesting avenues for future exploration. 

Our work also raises the question as to whether generic models of atomistic active matter such as Active Brownian or Run-and-Tumble particles \cite{BechingerVolpe2016} in which AFTs purportedly model \cite{Cates2019} are truly hyperuniform \cite{StenhammarCates2014}. Or that perhaps, while $\xi_h / l_1 \sim \mathcal{O}(0)$ for AFTs, fluctuations in agent-based models only become anomalously suppressed at much larger length scales beyond observational limits of current simulations, as indicated in \cite{LeiCiamarraNi2019}. Our study thus calls for greater understanding of $\xi_h/l_1$ for various existing continuum and atomistic active models, especially at time scales where the transition to anomalous fluctuation suppression remains observable. Lastly, higher-order correlations that lie beyond what is indicated by two-point functions exist. These correlations manifest as deviations in the higher central moments of $P(\phi_R)$, and persists at $R$ for which the AFTs are already indistinguishably fluctuation suppressed.  {Our results thus provide a pathway to identify correspondences between \emph{predictions} of particular AFTs and the underlying microscopic models they purportedly represent, or towards inferring the presence of novel interactions in experiments where the microscopic behavior is unknown}.

\begin{acknowledgments} 
  YZ acknowledges support form the Alexander-von-Humboldt
  Stiftung. The work of MAK and HL was supported by the SPP 2265 within the project LO 418/25. MAK also acknowledges support via the project KL 3391/2 and by the Initiative and Networking Fund of the Helmholtz Association through the Project ``DataMat''. 
\end{acknowledgments}

\end{document}